\documentclass[prc,aps,twocolumn,nofootinbib]{revtex4-1}
\usepackage[utf8]{inputenc}
\usepackage{amsmath}
\usepackage{amsfonts}
\usepackage{amssymb}
\usepackage{graphicx}
\usepackage{color}
\usepackage{hyperref}
\usepackage{xspace}
\usepackage{dcolumn}
\usepackage{cleveref}

\newcommand{\figbreak}{\newline}

\newcommand{\HS}{Hagedorn state\xspace}
\newcommand{\HSS}{Hagedorn states\xspace}

\newcommand{\dd}{{\rm d}}

\newcommand{\ee}{{\rm e}}

\newcommand{\UNIT}[1]{\ensuremath{\,{\rm #1}}\xspace}

\newcommand{\MeV}{\UNIT{MeV}}
\newcommand{\GeV}{\UNIT{GeV}}

\newcommand{\AGeV}{\UNIT{AGeV}}

\newcommand{\fm}{\UNIT{fm}}
\newcommand{\mb}{\UNIT{mb}}
\newcommand{\proz}{\UNIT{\%}}

\newcommand{\clebsch}[6]{\left\langle  #1\, #2 ; #3\, #4 \vert #5\, #6 \right\rangle^{2}}

\definecolor{magenta}{cmyk}{0,1,0,0}

\newcommand{\REM}[1]{}

\begin{document}

\title{Strangeness Production in low energy Heavy Ion Collisions via
  Hagedorn Resonances}

\author{K.~Gallmeister}
\author{M.~Beitel}
\author{C.~Greiner}
\affiliation{Institut f\"ur Theoretische Physik, Goethe-Universit\"at
  Frankfurt am Main, Max-von-Laue-Str.~1, 60438 Frankfurt am Main,
  Germany}

\begin{abstract}

  \begin{description}
  \item[Background] Statistical models are successfully used to describe particle multiplicities in (ultra-)relativistic heavy ion collisions. Transport models usually lack to describe special aspects of the results of these experiments, as the fast equilibration and some multiplicity ratios.

  \item[Purpose] A novel, unorthodox picture of the dynamics of heavy ion collisions is developed using the concept of Hagedorn states.

  \item[Method] A prescription of the bootstrap of Hagedorn states respecting the conserved quantum numbers baryon number $B$, strangeness $S$, isospin $I$ is implememted into the GiBUU transport model.

  \item[Results] Using a strangeness saturation suppression factor suitable for nucleon-nucleon-collisions, recent experimental data for the strangeness production by the HADES collaboration in Au+Au and Ar+KCl is reasonable well described. The experimental observed exponential slopes of the energy distributions are nicely reproduced.

  \item[Conclusions] A dynamical model using Hagedorn resonance states, supplemented by a strangeness saturation suppression factor, is able to explain essential features (multiplicities, exponential slope) of experimental data for strangeness production in nucleus-nucleus collisions close to threshold.

  \end{description}

\end{abstract}

\maketitle

\section{Introduction}
\label{sec:Introduction}

Statistical models are considered to be a valuable tool to understand the properties of the matter generated in (ultra-)relativistic heavy ion collisions \cite{BraunMunzinger:2003zd,Andronic:2008gu,Becattini:2000jw}, but also in high energetic $e^+e^-$,~$pp$ and $p\overline{p}$ collisions \cite{Becattini:1997rv}.

On the other hand, starting with Fermi's statistical model \cite{Fermi:1950jd}, fireball models were applied successfully for the description of experimental data in nucleon-antinucleon annihilation events at low energies. With the invention of the statistical bootstrap model (SBM) by Hagedorn \cite{Hagedorn:1965st} and its microcanonical reformulation by Frautschi \cite{Frautschi:1971ij} a systematical and consistent way for the inclusion of higher mass resonances was formulated. Hamer was the first to apply the (microcanonical) SBM to nucleon-antinucleon collisions \cite{Hamer:1972wz} (see also \cite{Dover:1992vj,Klempt:2005pp} for recent reviews).

In refs.~\cite{Beitel:2014kza,Beitel:2016ghw, Beitel2017} the authors developed a prescription of a microcanonical bootstrap of Hagedorn states with the explicitly conserved quantum numbers baryon number $B$, strangeness $S$ and charge $Q$. The covariant formulation is analogous to \cite{Hamer:1972wz,Yellin:1973nj}. Due to the restriction to only two constituents, a simple picture of creation and decay of \HSS as $2\to1$ and $1\to2$ processes could be formulated on the basis of detailed balance. This allows for dynamical calculations, and the model was successfully implemented into the hadronic transport model UrQMD\cite{Bass:1998ca,Bleicher:1999xi}. It has been shown in a first step, that from the Hagedorn model alone hadron multiplicities from cascading decay chains of a single heavy \HS are close to experimental data and that the energy spectra of the resulting hadrons from these decays follow an exponential law and thus look thermal by itself. Secondly, by performing box calculations, a desired fast equilibration time of strange and multistrange baryons and mesons was extracted.

Unfortunately, the numerical effort for these processes is quite high and especially the calculation of the decay of a \HS becomes slow when the mass of the resonance increases.
An alternative prescription using isospin, $I$, instead of charge, $Q$, is presented here and allows for much faster calculations. Therefore, for the first time, dynamical calculations of heavy ion collisions become feasible. Here now the transport model GiBUU \cite{Buss:2011mx} is used. This presents a new, unorthodox picture of the microscopical processes of the dynamics of heavy ion collisions. It has to be confronted against experimental findings and also against results of traditional transport calculations. In this pioneering and exploratory work the production of $\phi$ mesons in low energetic heavy ion collisions will be used as the test ground.

Recently, the HADES Collaboration has studied the role of the $\phi$ meson for the production of $K^-$ mesons in Ar+KCl collisions at a kinetic beam energy of 1.756\AGeV \cite{Agakishiev:2009ar} and in Au+Au collisions at 1.23\AGeV \cite{AdamczewskiMusch:2017rtf,SchuldesPhD}, covering the region $\sqrt{s_{NN}}=2.4-2.6\GeV$. Their data are compatible with former results by other experiments. The measured $\phi/K^-$ ratio was found to be $0.4-0.5$, meaning that $\approx 18-25\proz$ of the $K^-$ originate from decays of $\phi$ mesons. The spectra of the produced particles are thermal with slope parameters $T_{\rm eff}=70-100\MeV$. Traditional transport models have problems explaining theses findings.

It has to be noted that the idea of using higher mass resonances as a possible explanation of the above mentioned HADES data was already used in ref.~\cite{Steinheimer:2015sha}. Contrary to that work, the present approach represents a consistent way for the introduction of higher mass resonances.

The transport model GiBUU has been used for a long time to study low energetic heavy ion collisions \cite{Lang:1990zc} and strangeness production therein \cite{Wagner:2004ee,Larionov:2007hy}. It was also used for understanding production of hypernuclei in heavy ion collisions and antiproton induced reactions (see \cite{Gaitanos:2016pjg} for a recent survey). The interaction of the $\phi$ meson with hadronic matter has been studied in the case of photoabsorption on nuclei \cite{Muhlich:2002tu,Muhlich:2005kf}. In addition, the hadronic resonance model of GiBUU has been tested against dilepton measurements of many experiments, especially against HADES measurements of proton induced events, C+C collisions at 1 and 2\AGeV and the (above mentioned) Ar+KCl at 1.76\AGeV \cite{Weil:2012yg,Weil:2012ji}. It has to be mentioned, that the prescription of Kaon production in p+p collisions in GiBUU has recently been improved to match experimental data of the HADES group \cite{Agakishiev:2014moo}.

The paper is organized as follows:
In \cref{sec:Model} the mayor equations for the actual bootstrap model are shortly given and the overlap and the differences to previous work is indicated. The implementation of a strangeness saturation suppression factor is described.
Then, in \cref{sec:HeavyIonCollisions}, results from the present calculations are compared with experimental data from the HADES collaboration for A+A collisions. In a first step, only total particle multiplicities are considered, while in a second step, also energy spectra of particles are compared.
Finally, conclusions are drawn in \cref{sec:Conclusions}.

\section{Model}
\label{sec:Model}

Hagedorn states are hadron-like resonances, which are not limited to quantum numbers of known hadrons and also can be much heavier than known resonances. In the presented approach, these states are characterized by the quantum numbers baryon number ($B$), strangeness ($S$) and isospin ($I$). Instead of the latter, also the electrical charge ($Q$) could be used.

In the following a microscopic and dynamic description is provided of how a hadron resonance gas could be consistently expanded by Hagedorn states. Hagedorn state creation from two hadrons, their interaction with hadrons and other Hagedorn states and finally their decay into hadrons and/or other Hagedorn states are developed in a microcanonical way by respecting all the above given quantum numbers explicitly in each step. All this can be implemented into the transport model GiBUU\cite{Buss:2011mx}, replacing most of its default interactions.
Also, the resulting mass degeneracies can be used to enrich thermal model prescriptions by Hagedorn states, as in \cite{NoronhaHostler:2008ju,Majumder:2010ik}. Actually, the used prescription is a close extension of \cite{Belkacem:1998gy,Bratkovskaya:2000qy}.

\subsection{Basic equations}

The first mayor equation for the microcanonical bootstrap model with some conserved quantum numbers $\vec C$ \cite{Beitel:2014kza,Beitel:2016ghw, Beitel2017} is the bootstrap equation,\footnote{A factor $1/2!$ was missing in \cite{Beitel:2014kza,Beitel:2016ghw}.}
\begin{align}
\tau_{\vec C}(m)&=\tau^0_{\vec C}(m)+\frac{V(m)}{(2\pi)^2}\,\frac1{2m}
\ \sideset{}{^*}\sum_{\vec C_1,\vec C_2}\iint\dd m_1\dd m_2\nonumber\\
&\,\tau_{\vec C_1}(m_1)\tau_{\vec C_2}(m_2)
\,m_1\,m_2\,p_{\rm cm}(m,m_1,m_2)\quad,
\label{eq:basic1}
\end{align}
which tells, how the mass degeneration spectrum of the \HSS $\tau_{\vec C}(m)$ is build up from a low mass input $\tau^0_{\vec C}(m)$ and the combination of two lower lying \HSS. Here, $\tau^0_{\vec C}(m)$ may be identified with the spectral functions (delta function or Breit-Wigner) of the input hadrons.

The second fundamental equation is the connection between decay width and production cross section,
\begin{align}
\Gamma_{\vec C}(m)&=\frac{\sigma(m)}{(2\pi)^2}\,\frac1{\tau_{\vec C}(m)-\tau^0_{\vec C}(m)}
\ \sideset{}{^*}\sum_{\vec C_1,\vec C_2}\iint\dd m_1\dd m_2\nonumber\\
&\,\tau_{\vec C_1}(m_1)\tau_{\vec C_2}(m_2)
\,p_{\rm cm}^2(m,m_1,m_2)\quad.
\label{eq:basic2}
\end{align}
Here, the production cross section, $\sigma = \pi R^2$, and the volume of the Hagedorn resonances, $V=\frac43\pi R^3$, are given by a single radius parameter, $R$.
In a more general picture, the radius parameter $R$ could be taken mass dependent, $R=R(m)$,  $\sigma$ and $V$
The center of mass momentum is given as usually as
\begin{align}
p_{\rm cm}^2(m,m_1,m_2)&=\frac{(m^2-m_1^2-m_2^2)^2-4m_1^2m_2^2}{4m^2}\quad.
\end{align}
We introduced the notation $\vec C$ for the set of conserved quantum numbers and as abbreviation
\begin{align}
\sideset{}{^*}\sum_{\vec C_1,\vec C_2} =
\sum_{\vec C_1,\vec C_2}\delta(\vec C;\,\vec C_1,\vec C_2)
\end{align}
with
\begin{align}
\delta(\vec C;\,\vec C_1,\vec C_2)
=\delta(C^a;\,C_1^a,C_2^a)\,\delta(C^b;\,C_1^b,C_2^b)\cdots
\label{eq:generalDelta}
\end{align}
for indicating, that the summation just runs over the quantum number combinations, which are compatible with the overall quantum numbers.

In the case of additive discrete quantum numbers, as e.g.~baryon number $B$, strangeness $S$, and charge $Q$, the ``generalized'' delta symbol $\delta(z;x,y)$ in \cref{eq:generalDelta} is the usual one,
\begin{align}
\delta(X;X_1,X_2)&=\delta_{X,X_1+X_2}\ \text{for}\ X=B,\,S,\,Q,\,\dots\quad.
\end{align}
This set of quantum numbers ($B,S,Q$) was used in \cite{Beitel:2014kza,Beitel:2016ghw}, where the model was implemented into the hadronic transport model UrQMD \cite{Bass:1998ca,Bleicher:1999xi}.
This implementation has been shown to be rather ineffective and slow in computation time. Therefore the set of quantum numbers ($B,S,I$) with $I$ standing for the isospin will be used in present work.

The Gell-Mann--Nishijima formula ($2 I_z = 2Q-B-S$) connects the charge $Q$ (for known $B$ and $S$) with the $z$-component of the isospin, $I_z$, not with the isospin, $I$, directly. Therefore, the corresponding Clebsch Gordan coefficients have to be respected in the ``generalized'' $\delta$-symbol in \cref{eq:generalDelta},
\begin{align}
\delta(I;I_1,I_2)&=\begin{cases}1&\exists I^z,I_1^z,I_2^z: \clebsch{I_1}{I_1^z}{I_2}{I_2^z}{I}{I^z}\neq 0\\0&\text{otherwise} \end{cases}\quad .
\end{align}
While thus the equations look more difficult, the actual calculation is much faster, because the number of possible quantum number combinations to consider is much smaller.

It has been tested, that both approaches, i.e.~the new ($B,S,I$) and the former ($B,S,Q$) approach, give the same results, when the quantum numbers are identical. This needs some rework, since if the quantum numbers are fixed e.g.~to some values of ($B,S,Q$), one has to iterate in the ($B,S,I$) approach over all states, which may contribute to the given state. From the Gell-Mann--Nishijima formula we get some $I_z$ value, which indicates the minimal $I$ value of the iteration.

The presented prescription preserves quantum numbers in a microcanonical sense such that some quantum numbers are conserved explicitly. Unfortunately, some other quantities are not preserved. So the G-parity is violated; processes like $2\pi\rightarrow X\rightarrow 3\pi$ are possible. This may have direct consequences on the number of produced particles. Such considerations are left for further studies.

In the large mass region ($m\gtrsim3\GeV$), the resulting mass degeneracy can be fitted very well with a function including an exponential increase,
\begin{align}
\tau_{\vec C}(m)\stackrel{m\to\infty}{\longrightarrow}a_{\vec C}\ m^{-b_{\vec C}}\ \ee^{c_{\vec C}\,m}\quad,
\label{eq:fitfunction}
\end{align}
were the three parameters $a$, $b$, and $c$ depend on the quantum numbers $\vec C$. The parameter $c$ depends only very weakly \cite{Beitel:2014kza,Beitel:2016ghw, Beitel2017}; it is assigned with the label ``Hagedorn temperature'', $T_{H,\vec C}\equiv 1/c_{\vec C}$. Usually, this notion is connected with the temperature of the system, where the partition function diverges. Since the present model is constraint in the bootstrap to some maximal mass ($m<10\GeV$) for the \HSS due to numerical reasons, the actual divergence can not fully been observed. Therefore only an approximate value of the Hagedorn temperature derived from the fits according \cref{eq:fitfunction} can be given, $T_H = \langle T_{H,\vec C}\rangle$, were the averaging is done over all possible quantum number states $\vec C$.

The only free parameter of the presented statistical bootstrap is the radius parameter $R$, which enters \cref{eq:basic1,eq:basic2} as the volume $V$ of the \HS and as the production cross section $\sigma$. But it also directly influences the slopes of the spectra and is thus directly connected with the value of the Hagedorn temperature $T_H$. In the present work, a fixed value $R=1.0\fm$ is chosen, yielding $\sigma=31\mb$, $V=4.2\fm^3$, and a Hagedorn temperature $T_H\sim165\MeV$. Due to numerical reasons, we have to restrict to masses $m<10\GeV$ in the bootstrap.

\subsection{Phase space diagram}

\begin{figure}[htb]
  \begin{center}
    \hspace*{\fill}%
    \includegraphics[width = 0.45\textwidth,clip=true]{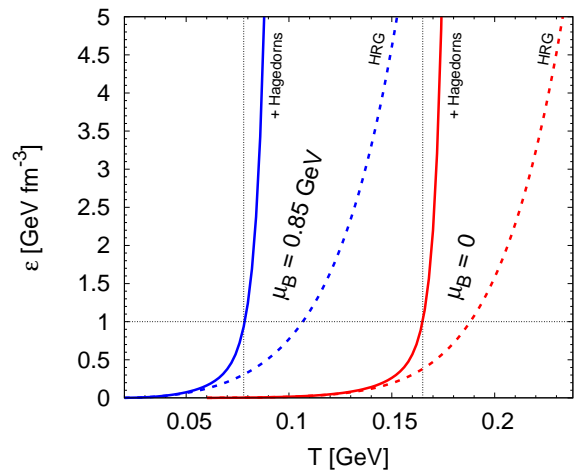}
    \hspace*{\fill}%

    \caption
    {\textit{The energy density as function of the temperature for $\mu_B=0$ (red curves) and $\mu_B=0.85\GeV$ (blue curves) with vanishing net strangeness density. Solid curves show results including Hagedorn states, dashed curves a hadron gas only.
      }}
    \label{fig:EvsT}
  \end{center}
\end{figure}

As in \cite{Belkacem:1998gy,Bratkovskaya:2000qy}, the calculated Hagedorn spectra may be included into a statistical model with baryo-chemical potential $\mu_B$ and strange potential $\mu_S$.
As mentioned above one normally connects the notion of a Hagedorn temperature with that temperature, where the partition function diverges. In order to illustrate this behavior, the divergence of the energy density as function of the temperature is shown in \cref{fig:EvsT}.
Here results for vanishing baryo-chemical potential, $\mu_B=0$, are compared to those with some non-vanishing value. For the latter case, the chemical potential $\mu_S$ is adjusted to guarantee vanishing net strangeness. (Resulting curves with fixed $\mu_S=0$ vary just within temperatures of $1-2\MeV$.) While for vanishing baryo-chemical potential the curve starts to increase very rapidly close to the fitted $T_H$ above, the divergence occurs for large values of $\mu_B$ at clearly smaller temperatures. Again, if in the present model Hagedorn masses would not be limited, the divergence would be much sharper.

\begin{figure}[htb]
  \begin{center}
    \hspace*{\fill}%
    \includegraphics[width = 0.45\textwidth,clip=true]{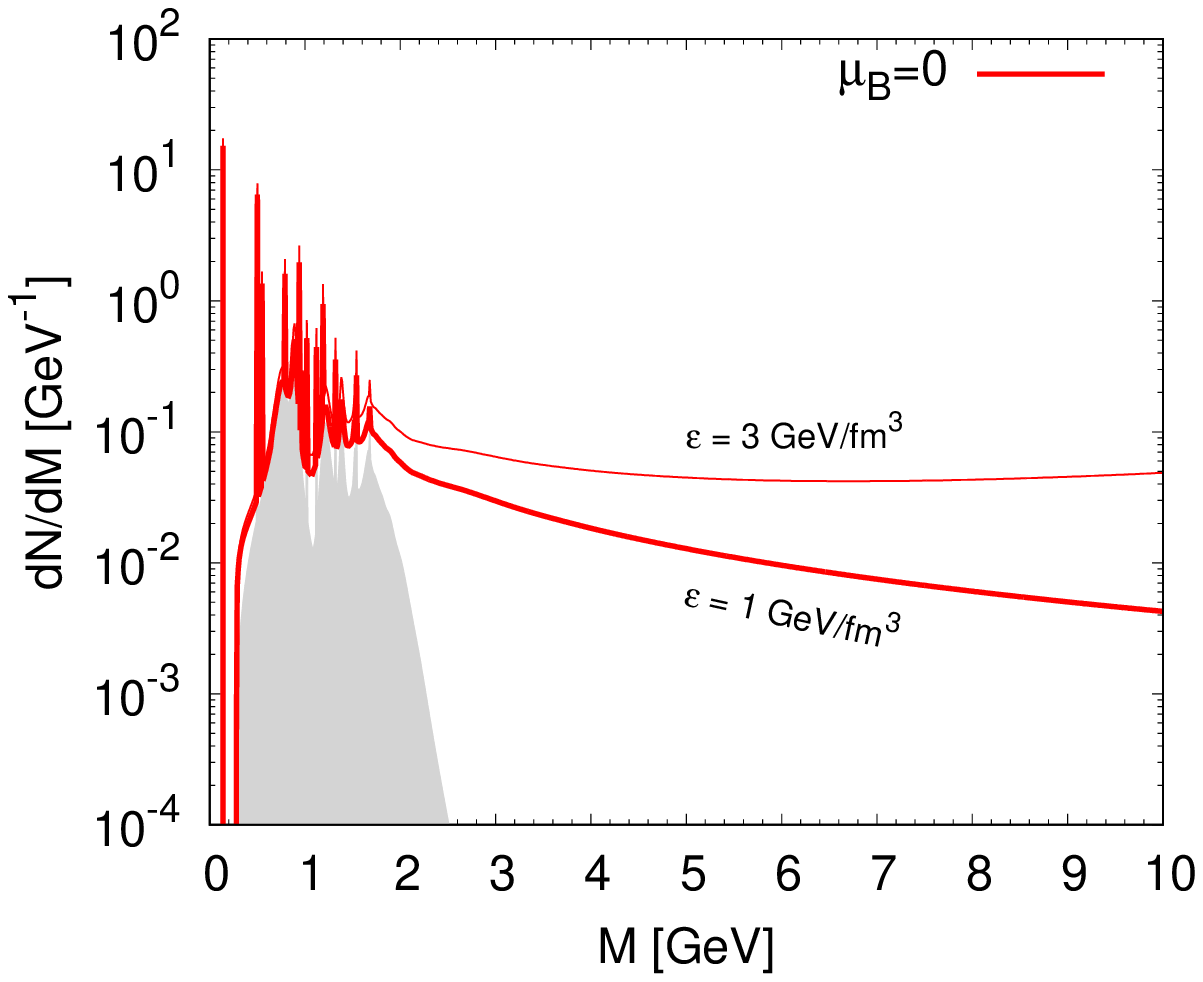}
    \hspace*{\fill}%

    \caption
    {\textit{The mass distribution for vanishing baryo-chemical potential. The gray area shows the hadronic contribution, while the lines indicate the results for given $\epsilon$.
      }}
    \label{fig:EpsMass0}

    \hspace*{\fill}%
    \includegraphics[width = 0.45\textwidth,clip=true]{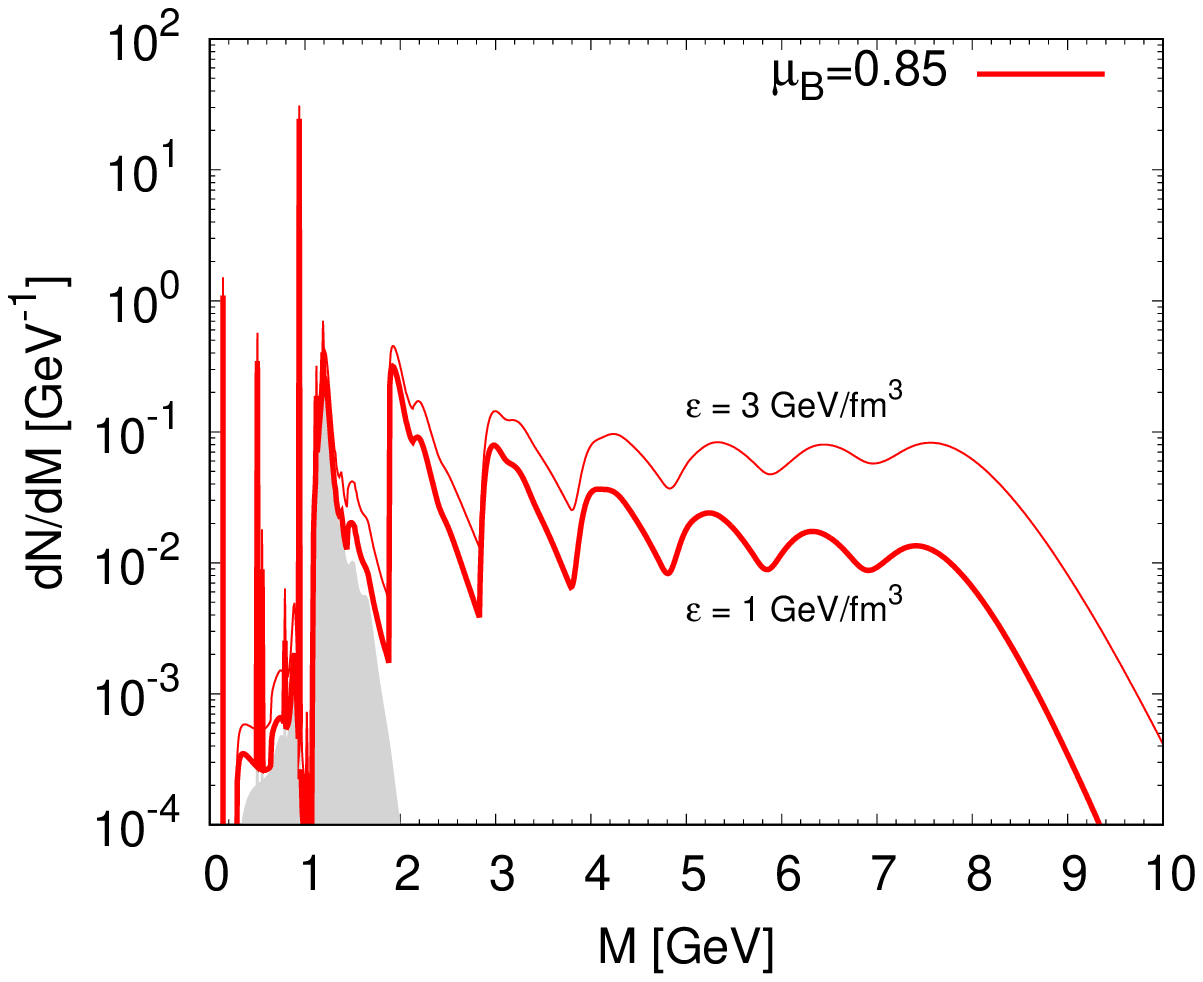}
    \hspace*{\fill}%

    \caption
    {\textit{As \cref{fig:EpsMass0}, but for $\mu_B=0.85\GeV$.
      }}
    \label{fig:EpsMass1}
  \end{center}
\end{figure}

For large temperatures $\sim T_H$, the mass spectrum for every quantum number gets enhanced for large masses. This is shown in \cref{fig:EpsMass0}, where the mass spectrum yielding energy densities $\epsilon=1-3\GeV\fm^{-3}$ are shown.
At least for the latter condition the spectrum seems to increase with increasing mass, yielding definitely a diverging partition sum.

For the case of a large value of the baryo-chemical potential $\mu_B$, the situation is different. The connected temperature leading to a divergence is much smaller ($\sim75\MeV$, see \cref{fig:EvsT}), yielding slopes, which drop very fast with increasing mass, when the net baryon number is fixed. The large factor $\ee^{B\,\mu_B/T}$ enhances more and more particles with larger baryon numbers. Finally, in the limit of vanishing temperature but maximal baryo-chemical potential, the mass distribution would consist of a sum of equipotent, rather spiky functions, located at multiples of the nucleon mass $m_N=0.938\GeV$. This is clearly and prominently visible in \cref{fig:EpsMass1}. Obviously, only particles with $B\leq8$ are considered in the present prescription.
Nevertheless, this constraint does not influence the results. E.g., the inclusion of baryon numbers reduced by 1 only shifts the divergence region by $\sim2\MeV$.

In the case of vanishing baryo-chemical potential, the ratio of particles with strangeness $|S|=1$ is $35-45\proz$ for masses $M>2\GeV$. The content of multistrange particles increases with increasing mass and reaches 25\proz, 15\proz, and 10\proz for $|S|=2$, $|S|=3$ and $|S|>3$ at $M=10\GeV$. This ratio is independent of the underlying energy density.
For the case of $\mu_B=0.85\GeV$, the two different treatments of the strange chemical potential yield different contributions. In the case of $\mu_S=0$, the strangeness content oscillates with mass according multiples of $m_N$ and reaches approximately the same values as above in maximum. In between, the strange contribution drops close to zero. If $\mu_S$ is varied in order to guarantee a vanishing net strangeness, the total strangeness is significantly suppressed. Only approximately 20\proz of all particles have strangeness $|S|=1$. Multistrange states with $|S|>1$ do not play any role at all.

\begin{figure}[htb]
  \begin{center}
    \hspace*{\fill}%
    \includegraphics[width = 0.45\textwidth,clip=true]{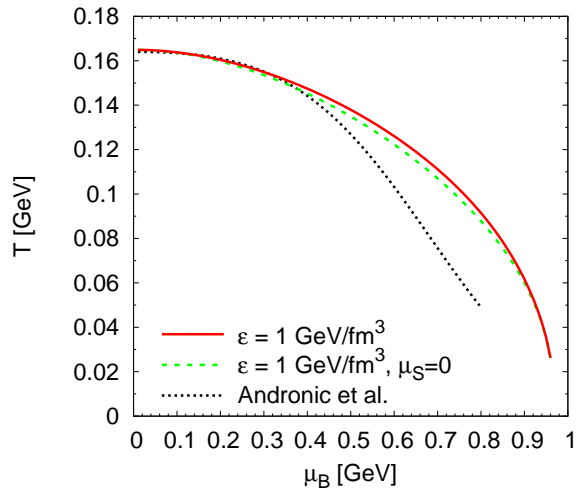}
    \hspace*{\fill}%

    \caption
    {\textit{The boundary $\epsilon=1\GeV\fm^{-3}$ in the $T-\mu$ plane. Also shown the freeze-out curve from \cite{Andronic:2005yp,Andronic:2008gu}.
      }}
    \label{fig:Eps1Curve}
  \end{center}
\end{figure}

It is instructive to study the divergence in the $T-\mu$ plane. Since it is not possible to show the real divergence, some cut into the energy distribution may serve as a hint to the phase boundary. This is done in \cref{fig:Eps1Curve}, where the value $\epsilon=1\GeV\fm^{-3}$ is chosen.
Low temperatures ($T<25\MeV$) are excluded from this consideration, since here the Bose- or Fermi-nature of the particles become important.
As also shown in this figure, the treatment of the strange chemical potential is only of minor importance for this boundary. In addition, the boundary is compared with the freeze-out parametrization from \cite{Andronic:2005yp,Andronic:2008gu}. While for small values of the baryo-chemical potential, the two curves lie on top of each other, the boundary of the Hagedorn scenario reaches larger values of $\mu_B$ for a given temperature.

\subsection{Strangeness saturation suppression factor and N+N collisions}

It is well known, that statistical models, including the SBM, yield too large multiplicities of strangeness carrying particles compared to data of the annihilation experiments \cite{Hamer:1972wz}(and references therein).
Dynamical suppression according the OZI rule \cite{Okubo:1963fa,*Zweig:1964jf,*Iizuka:1966fk} could be an explanation.

Therefore, in the following, a strangeness saturation suppression factor $\gamma_s$ is implemented by rescaling the cross section (and via the detailed balance constraint also the decay width) by a 'penalty' factor $\gamma_s^2$, if the creation or deletion of a $s\overline s$ quark pair is involved. This can be formulated by the replacement
\begin{align}
\sigma(m)\longrightarrow \sigma(m)\,\gamma_s^{|S_1|+|S_2|-|S_1+S_2|}\quad,
\label{eq:rescaleSigma}
\end{align}
in \cref{eq:basic2}, where $S_{1,2}$ is the strangeness content of the two incoming/outgoing particles.
A closer inspection of this equation and \cref{eq:basic1,eq:basic2} shows, that (when the inhomogenity $\tau^0$ can be neglected) this rescaling of the cross section $\sigma$ is equivalent to a rescaling of the $\tau$ itselves,
\begin{align}
\tau_{\vec C}(m) \longrightarrow \tau_{\vec C}(m)\,\gamma_s^{|S|}\quad.
\label{eq:rescaleTau}
\end{align}
Note that this rescaling should happen after the bootstrap, but before calculating $\Gamma$.

Also processes deleting or producing a $\phi$ meson have to get some 'penalty' factor $\gamma_\phi$, since the $\phi$ meson has some $s\overline s$ quark content, while its overall strangeness is zero.
In the following, the simplest assumption, $\gamma_\phi=\gamma_s^2$,  will be choosen.
The inclusion of this factor has to be done independent of the treatment \cref{eq:rescaleSigma} resp.~\cref{eq:rescaleTau} in some additional, separate step.

A different approach would already rescale the $\tau_{\vec C}^0$ according the prescription \cref{eq:rescaleTau} before the bootstrap. It has been tested, that this approach would yield similar results like shown below for the case for A+A collisions. Nevertheless, the thermodynamical limit would yield a different asymptotic state than given here. Therefore this approach is not followed here.

It has to be noted, that the dynamical approach described here is close to the thermal description described in ref.~\cite{Becattini:1997rv}: a microcanonical suppression in addition to a dynamical strangeness suppression factor $\gamma_s$.

\begin{figure}[htb]
  \begin{center}
    \hspace*{\fill}%
    \includegraphics[width = 0.45\textwidth,clip=true]{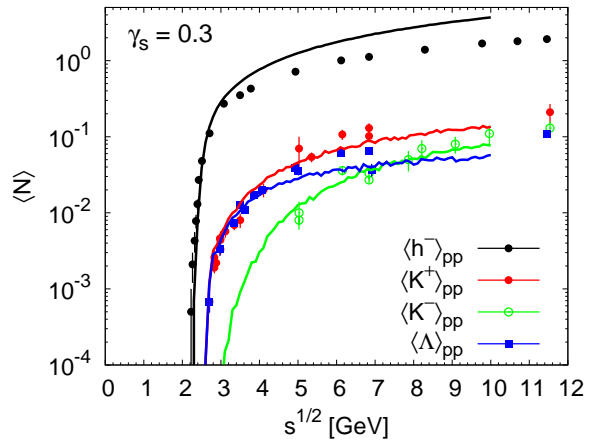}
    \hspace*{\fill}%

    \caption
    {\textit{The averaged number of negative charged hadrons $h^-$, $K$ mesons, and $\Lambda$ baryons from a Hagedorn state (B,S,I,(Q))=(2,0,1,(2)) decay compared with experimental data from p+p collisions as function of $\sqrt s$ (data compilation:\cite{Gazdzicki:1995zs,Gazdzicki:1996pk}). The strangeness suppression is $\gamma_s=0.3$.
      }}
    \label{fig:ppDat}
  \end{center}
\end{figure}

In order to adjust the value of $\gamma_s$, experimental data for strangeness production in low energetic p+p collisions is considered (data compilations \cite{Gazdzicki:1995zs,Gazdzicki:1996pk}). A comparison of calculations doing Monte Carlo decays of a \HS with the quantum numbers $(B,S,I)=(2,0,1)$ and the additional constraint $Q=2$ with the experimental data is shown in \cref{fig:ppDat}.
As a 'fit-by-eye', the value $\gamma_s=0.3$ will be used in this work.
This value is compatible with ref.~\cite{Becattini:1997rv}, where the energy dependence for energies larger than considered here shows a linear increase.

As can be seen in \cref{fig:ppDat}, the charged hadron (mainly pions) multiplicity is slightly underestimated directly at the threshold, but somewhat overestimated for $\sqrt s>3\GeV$. This behavior is hardly influenced by the choice of the $\gamma_s$ value. Also the excitation function of $K^+$ and $\Lambda+\Sigma^0$ is only described well in the overall view; in the details differences are visible. Unfortunately, data for $K^-$ multiplicities is not available in the region below 5\GeV, thus no conclusion about the quality of the theoretical prescription can be drawn.

The implemented strangeness suppression factor $\gamma_s$ is directly responsible for the multiplicity of the $K^+$ mesons and the $\Lambda$ and $\Sigma$ baryons. On the other hand, the yield of the $K^-$ mesons is connected with the yield of $\phi$ mesons. Here, the above mentioned factor $\gamma_\phi$ is responsible via the connection $\gamma_\phi=\gamma_s^2$.

It has been checked, that the description of the negatively charged hadrons in p+n  is similar to the shown case here; underestimation directly at the threshold, while for larger energies the data is overestimated.

As a note, another source of adjustment could be the large number of experimental observables, especially all the different final state channels in nucleon-antinucleon annihilation. Nevertheless, there the situation is much different, since due to the different total baryon number, the production channels for mesons are already open at threshold. Therefore these data are not really applicable to the calculation of heavy ion collisions, where also the baryon number plays a very import role. Thus the annihilation data will not be used here.

The A+A collisions discussed in this work cover the region $\sqrt{s_{NN}}=2.4-2.6\GeV$.

\section{A+A collisions}
\label{sec:HeavyIonCollisions}

In the following, heavy ion collisions are performed in the framework of GiBUU\cite{Buss:2011mx}. The colliding nuclei are initialized consisting of a given number of protons and neutrons. The nucleons have Fermi momentum and are bound in a potential. At initial time $t=0$, the nuclei are initialized with a necessary distance away from each other and then propagating onto each other. Centrality/impact parameter constraints are implemented according the experimental needs \cite{Agakishiev:2011zz,AdamczewskiMusch:2017rtf,SchuldesPhD}.
 When not stated otherwise, usual collisions are replaced by the formation of \HSS in this work.

Contrary to the usual prescription, \HSS can be formed by a first N+N collision, but then also gather additional energy by picking up a second nucleon. Thus, the \HSS can act as some kind of 'energy reservoir'.

\begin{figure}[htb]
  \begin{center}
    \hspace*{\fill}%
    \includegraphics[width = 0.45\textwidth,clip=true]{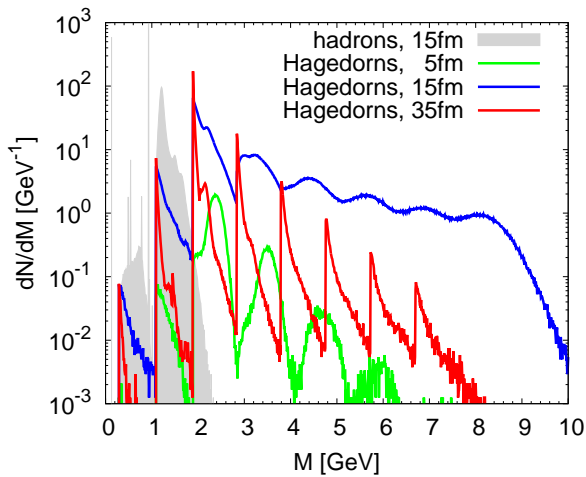}
    \hspace*{\fill}%

    \caption
    {\textit{The mass distribution of hadrons and \HSS at different time steps of the calculation for Au(1.23\AGeV)Au (0-40\proz) collisions. The maximal overlap of the nuclei is at $t=12.0\fm$.
      }}
    \label{fig:MassAuAu}
  \end{center}
\end{figure}

In order to illustrate this \HS scenario at work, the mass distribution of \HSS at a some fixed times for Au(1.23\AGeV)Au collisions is shown in \cref{fig:MassAuAu}.
The maximal overlap of the colliding nuclei happens at $t=12.0\fm$. Thus the maximal mass extend of the generated \HSS is approximately 3\fm after this point.

It is a remarkable feature, that very high mass states are populated. The distribution is very smooth in its maximal extend. Before and after this time, sharp structures are visible. At the beginning, \HSS are being built up by 2, 3, \dots{} nucleons. These nucleons have approximately  the same energy, therefore multiples of the initial kinetic energy governs the mass distribution of the \HSS. When decays of the \HSS set in, then the mass of the nucleons is the relevant scale. Then, again sharp structures in the mass distribution develop, but now different from the initial ones.

\begin{figure}[htb]
  \begin{center}
    \hspace*{\fill}%
    \includegraphics[width = 0.45\textwidth,clip=true]{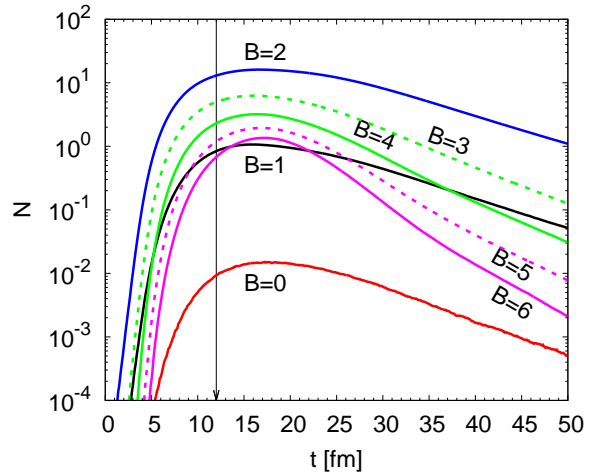}
    \hspace*{\fill}%

    \caption
    {\textit{The evolution of Hagedorn states (excluding known hadrons) with different baryon number $B$ for Au(1.23\AGeV)Au (0-40\proz) collisions. The maximal overlap of the nuclei is at $t=12.0\fm$. (The number of hadrons is too large to be visible in this plot.)
      }}
    \label{fig:BaryonAuAu}
  \end{center}
\end{figure}

The time evolution of states with different baryon numbers is illustrated in \cref{fig:BaryonAuAu}.
A similar picture as \cref{fig:BaryonAuAu} for different strangeness states is not very illustrative, since the number of states with $S=|1|$ is already suppressed by a factor of 1000, particles with larger strangeness do essentially not play any role at all.

\subsection{Multiplicites}

The transport model GiBUU features two different kinds of calculation setups. In the first one, called 'parallel ensemble', multiple ensembles are calculated in parallel, without any interference. The number of testparticles is unity; this setup is a microcanonical one. The other setup, called 'full ensemble', mixes all ensembles and thus the number of testparticles equals the number of ensembles. Therefore, this is a (grand)canonical setup. This is of importance for the production and absorption of rare particles \cite{Ko:2000vp,Fochler:2006et}.

\begin{figure}[htb]
  \begin{center}
    \hspace*{\fill}%
    \includegraphics[width = 0.45\textwidth,clip=true]{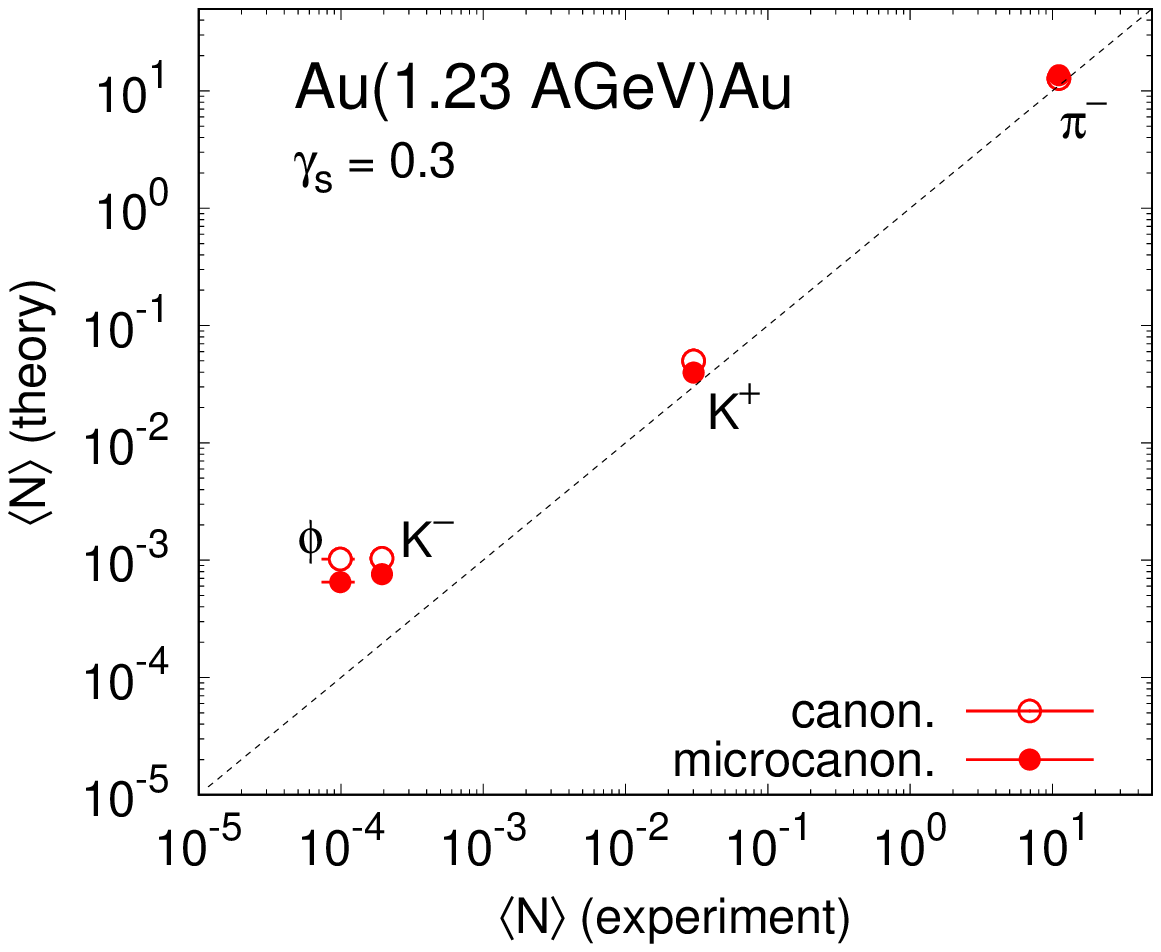}
    \hspace*{\fill}%
    \figbreak
    \hspace*{\fill}%
    \includegraphics[width = 0.45\textwidth,clip=true]{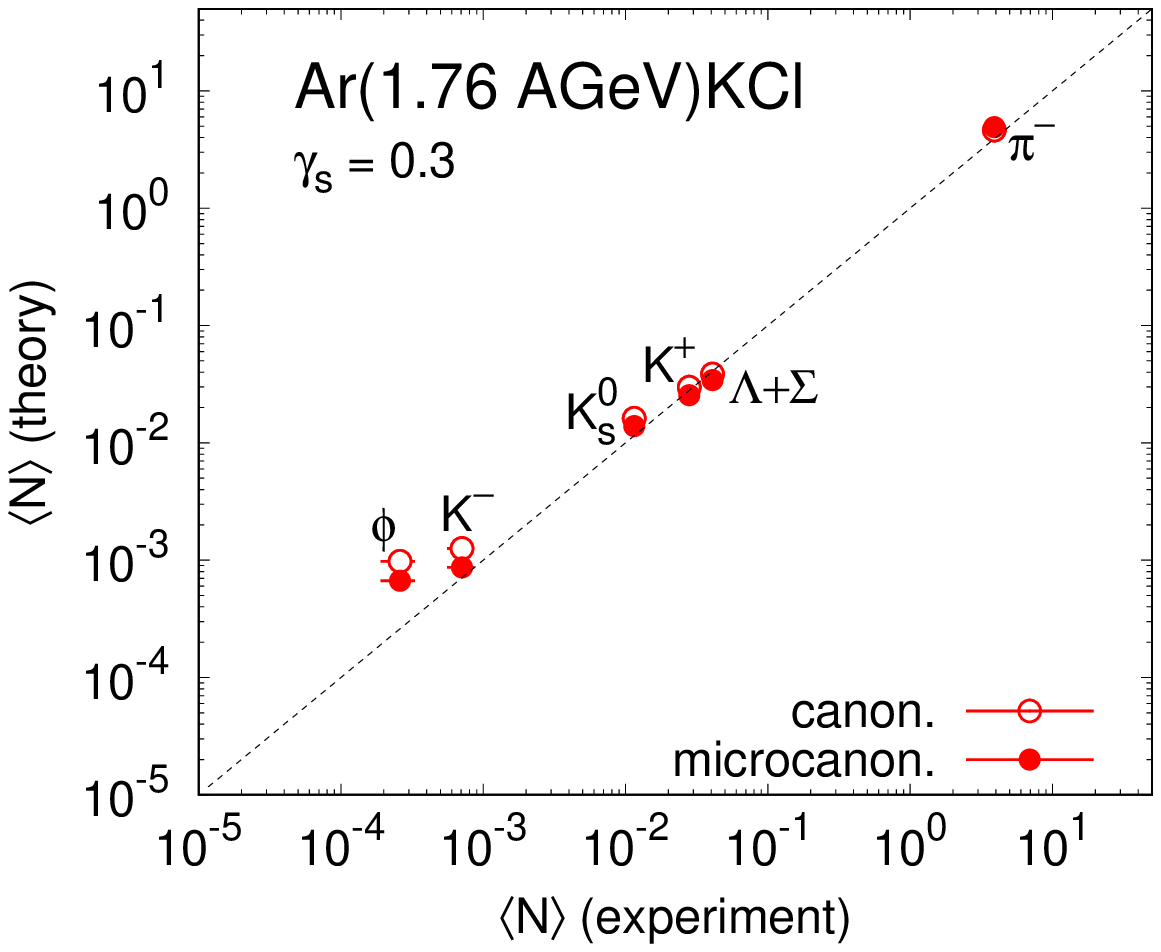}
    \hspace*{\fill}%

    \caption
    {\textit{Comparison of calculated and measured multiplicities in Au(1.23\AGeV)Au (0-40\proz) \cite{AdamczewskiMusch:2017rtf,SchuldesPhD} and Ar(1.76\AGeV)KCl (min.~bias) \cite{Agakishiev:2009ar}. The strangeness suppression is $\gamma_s=0.3$.
      }}
    \label{fig:mult}
  \end{center}
\end{figure}

In \cref{fig:mult}, the calculated multiplicities are compared with experimental data for  Au(1.23\AGeV)Au (0-40\proz) \cite{AdamczewskiMusch:2017rtf,SchuldesPhD} and Ar(1.76\AGeV)KCl (min.~bias) \cite{Agakishiev:2009ar}.
An effect of the microcanonical treatment is visible, since the multiplicities of the rare particles are even more suppressed than in the canonical treatment. Nevertheless, the $\phi$ multiplicities come out too high in the calculations, influencing the $K^-$ in the same manner. The effect is larger at the smaller energy. For the higher energy, the multiplicity of the $K^-$ is described very reasonable. The multiplicities of $\pi^-$, $K^+$, $K^0_s$ and $\Lambda + \Sigma^0$ are described well where measured.
The visible overestimation for the $\phi$ meson yields will be further discussed below when the spectra are considered.

One observes, that the $K^-$ are strongly produced via $\phi$ decays; the calculated $\phi/K^-$ ratio is in the order of $80\proz$ (cf.~\cref{{tab:phiK}}).
\begin{table}[htbp]
\begin{center}
\begin{ruledtabular}
\begin{tabular}{cr@{$\ \pm$}lr@{$\ \pm$}l}
&\multicolumn{2}{l}{Au(1.23\AGeV)Au}&\multicolumn{2}{l}{Ar(1.76\AGeV)KCl}\\
\hline
HADES & 0.52 & 0.16 & 0.37 & 0.13 \\
Hagedorn & 0.85 & 0.11 & 0.77 & 0.06 \\
GiBUU & 0.13 & 0.04 & 0.11 & 0.01 \\
\end{tabular}
\end{ruledtabular}
\caption{\textit{Values for the ratio $\phi/K^-$.
  }}
\label{tab:phiK}
\end{center}
\end{table}
It is important to keep in mind, that the fraction of $K^-$ stemming from $\phi$ decays is given as $K^-_{(\textrm{from }\phi)}/K^-=\phi/K^-\cdot{\cal B}_{\phi\to K^-K^+}$ with ${\cal B}_{\phi\to K^-K^+}$ giving the branching ratio for the decay of the $\phi$ meson into two charged kaons, which is taken as 0.42 in this model. Thus even with $\phi/K^-\sim 0.85$, only approximately 1/3 of the final $K^-$ stem from $\phi$ decays.

The $\phi$ mesons are only produced via Hagedorn resonance decays into a $\phi$ meson and a hadron/Hagedorn state, while approximately 33\proz of all produced $\phi$ is reabsorbed again. Here 60\proz of all interactions (production and reabsorption) are governed by $B=1$ Hagedorn states. All states with higher baryon number contribute, while $B=7$ is only weaker by 20\proz than $B=2$ (reabsorption factor is constant for all $B$). Comparing these numbers with \cref{fig:BaryonAuAu}, one finds a non-trivial dependency of $\phi$ production on the baryon number of the Hagedorn states.

\subsection{Traditional Transport (GiBUU)}

In order to digest the effects of the \HSS, the same calculations have been performed using the default interaction scenario implemented in GiBUU.
Here, no modifications for better descriptions of experimental date have been adopted. The equation of state etc.\ is kept as its default.
As mentioned above, GiBUU was earlier used to study strangeness production in more detail, see especially ref.~\cite{Agakishiev:2014moo}.

\begin{figure}[htb]
  \begin{center}
    \hspace*{\fill}%
    \includegraphics[width = 0.45\textwidth,clip=true]{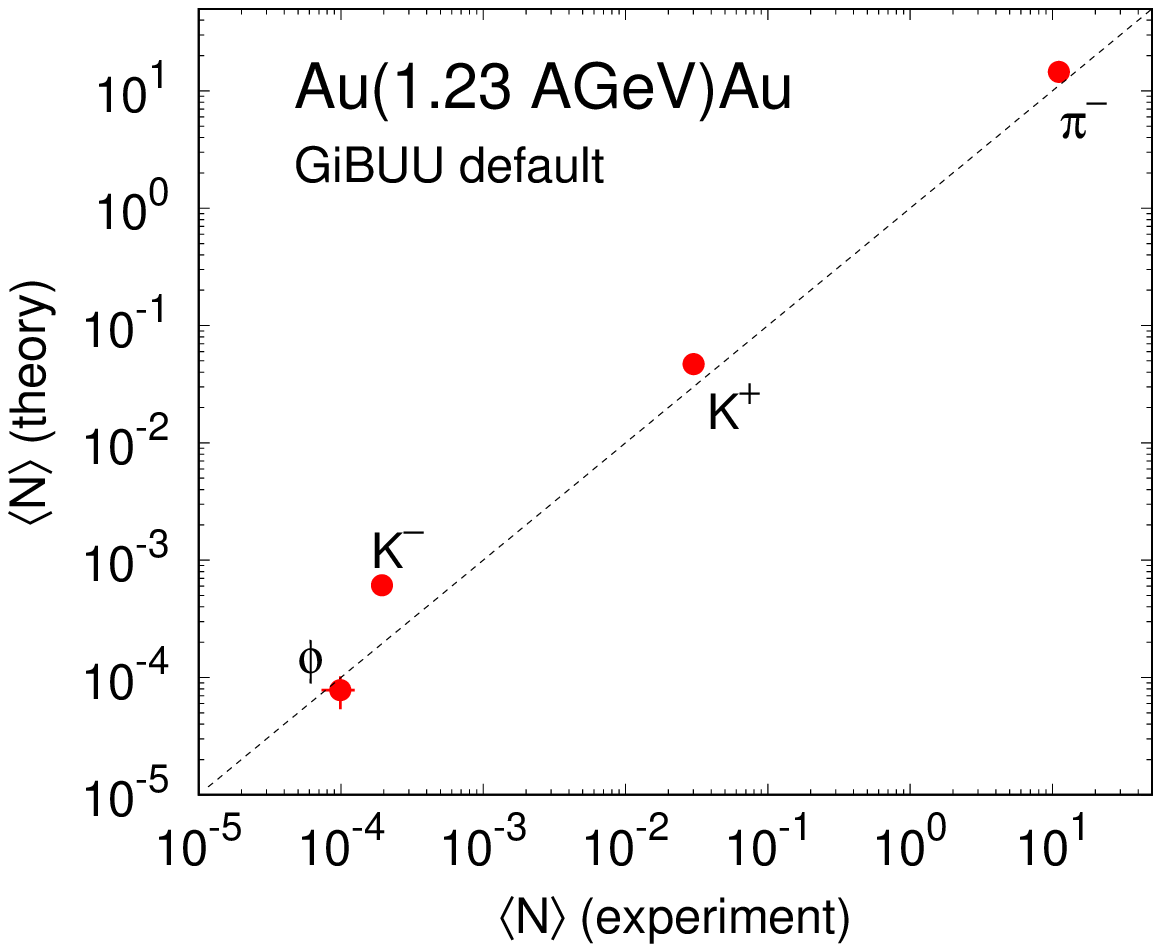}
    \hspace*{\fill}%
    \figbreak
    \hspace*{\fill}%
    \includegraphics[width = 0.45\textwidth,clip=true]{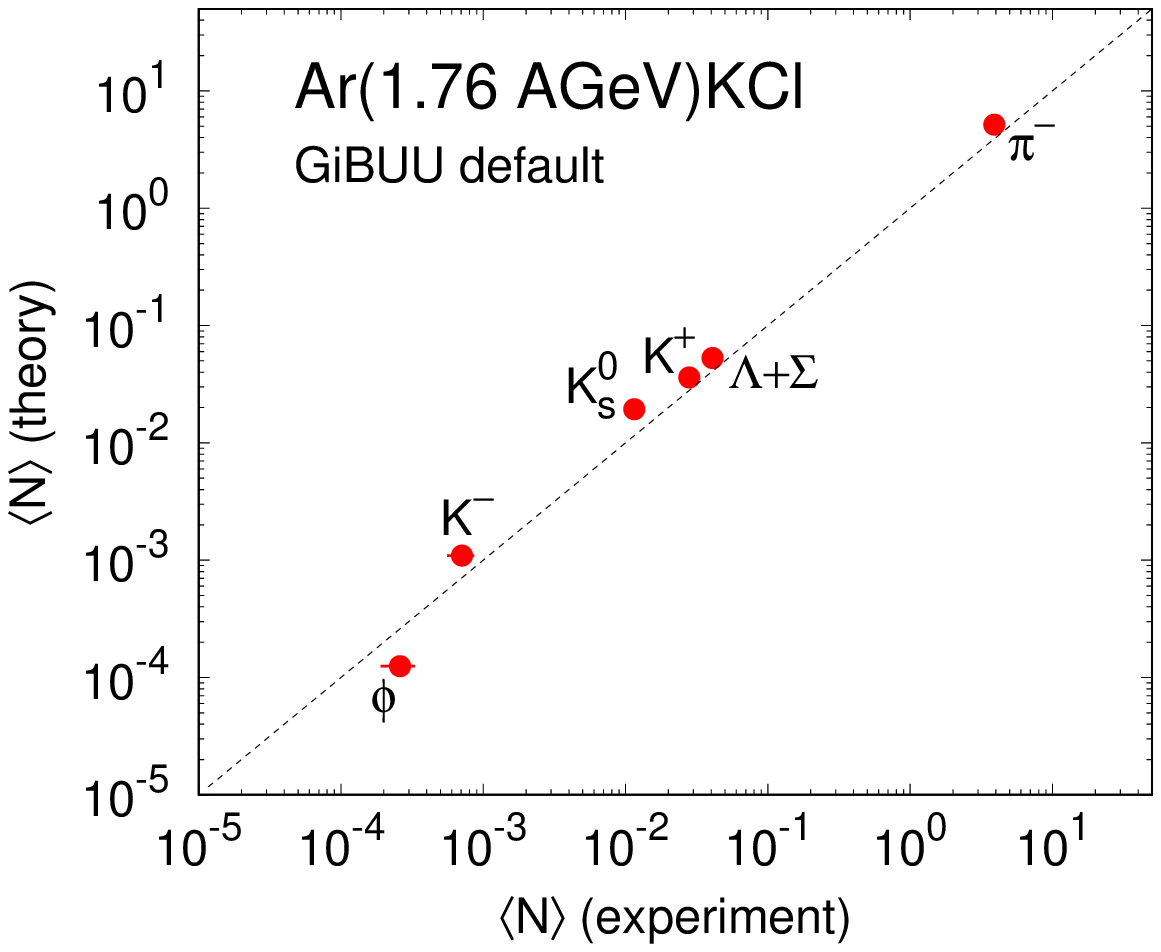}
    \hspace*{\fill}%

    \caption
    {\textit{Comparison of calculated and measured multiplicities in Au(1.23\AGeV)Au (0-40\proz) \cite{AdamczewskiMusch:2017rtf,SchuldesPhD} and Ar(1.76\AGeV)KCl (min.~bias) \cite{Agakishiev:2009ar}. GiBUU is used in its defaults; microcanonical prescription.
      }}
    \label{fig:multGiBUU}
  \end{center}
\end{figure}

The comparison of the resulting multiplicities with experimental ones is shown in \cref{fig:multGiBUU}.
The agreement between calculations and experiment is as good as within the Hagedorn approach. One significant difference is observable: In the default GiBUU approach, the $\phi/K^-$ ratio is much smaller ($\sim0.1$) than in the Hagedorn approach ($\sim0.8$), see \cref{tab:phiK}.
In addition, the $K^-$ multiplicities in the Hagedorn and in the GiBUU default approach are more or less identical, but only the origin of these mesons is different.
Thus, within the hadronic treatment of GiBUU, the $K^-$ are not mainly generated via $\phi$ decays.

In this pure hadronic treatment, approximately 60\proz of all produced $\phi$ mesons will be reabsorbed, mainly due to $N\phi\to N\pi\pi$ processes. For this process, the parametrizations by Golubeva et al.~\cite{Golubeva:1997na} are used in GiBUU, see e.g.~\cite{Muhlich:2002tu,Muhlich:2005kf}. At low energies, this cross section becomes larger than the (constant) cross section used in the Hagedorn picture. The main production mechanisms for $\phi$ production are the channels $\pi\rho\to\phi$ (35\proz) and $N\pi\to N\phi$ (20\proz).

\subsection{Slopes}

In addition to the multiplicities, also the spectra of the particles are examined. In the experiment, the data were divided into different rapidity bins, for which then the transverse mass spectra were fitted with some exponential slope\footnote{Actually, the fitting may be more involved. For details see \cite{AdamczewskiMusch:2017rtf,SchuldesPhD,Agakishiev:2009ar}.}. In a second step, the resulting temperatures as function of rapidity $y$ where fitted assuming a pure thermal source with a distribution $T_{\rm eff}/\cosh y$, yielding one final number. For the calculations, efficiency or acceptance considerations play no role and one can directly look at the energy spectra $\dd N/p E\dd E\sim \ee^{-E/T_{\rm eff}}$ (as long as one has not to expect any asymmetry of transversal and longitudinal direction).
Unfortunately, low numerical statistic do not allow for reasonable fit values for the theoretical curves.
The calculated spectra follow extremely well exponential slopes of a thermal source, except for the pions. Here some excess for low energies is given, which may be due to the contributions of pions stemming from hadronic decays and additional flow effects.

\begin{figure}[htb]
  \begin{center}
    \hspace*{\fill}%
    \includegraphics[width = 0.45\textwidth,clip=true]{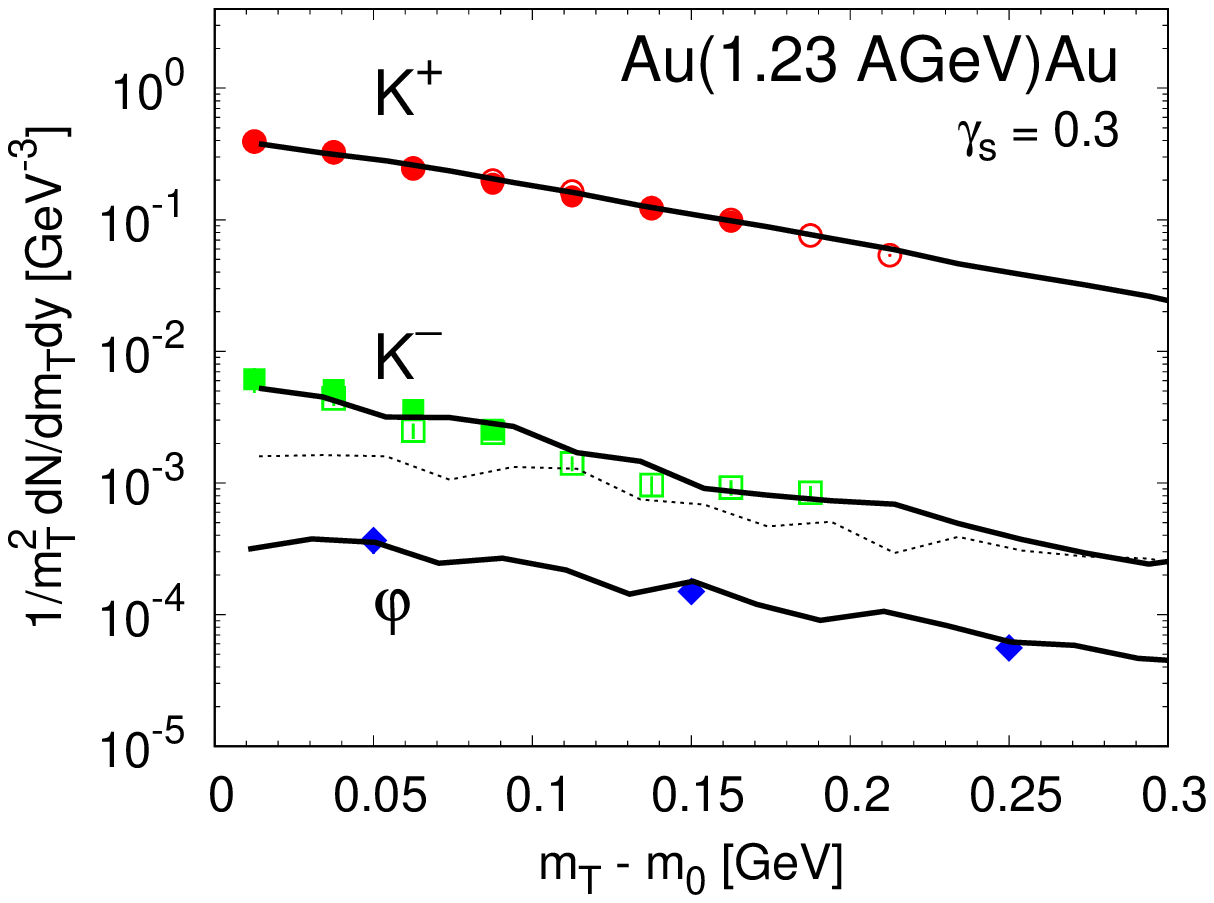}
    \hspace*{\fill}%

    \caption
    {\textit{Transverse mass spectra at midrapidity for Au(1.23\AGeV)Au (0-40\proz). Symbols show experimental data \cite{AdamczewskiMusch:2017rtf,SchuldesPhD} (solid: mid-rap, open: next-to-midrap). The dotted curve shows the $K^-$ spectrum without $\phi$ decays. Normalization of theoretical curves arbitrary.
      }}
    \label{fig:MTSAuAu}

    \hspace*{\fill}%
    \includegraphics[width = 0.45\textwidth,clip=true]{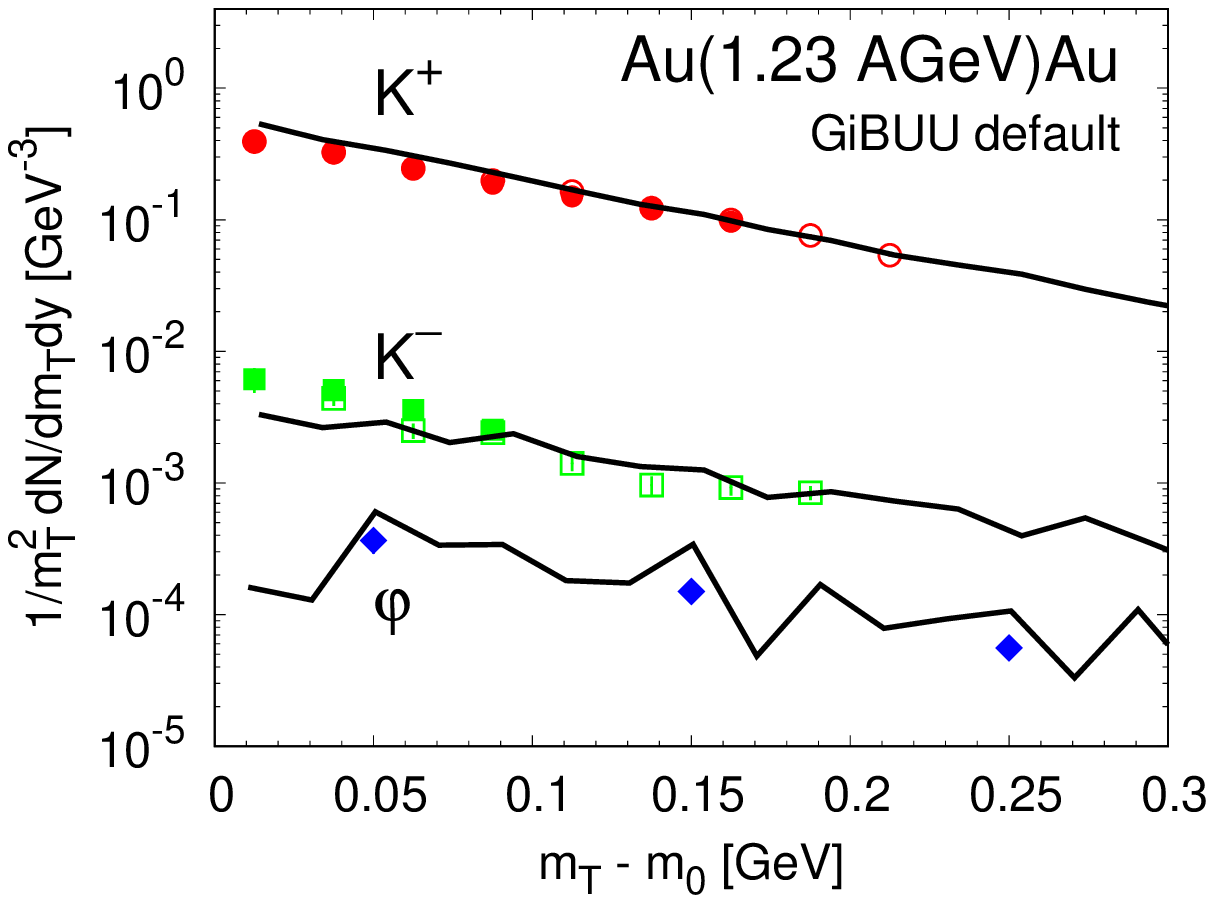}
    \hspace*{\fill}%

    \caption
    {\textit{As \cref{fig:MTSAuAu}, but with GiBUU in its defaults. ($\phi$ decay contribution to the $K^-$ spectrum not visible.)
      }}
    \label{fig:MTSAuAuHadron}

  \end{center}
\end{figure}

Thus, instead of comparing some fit values, a comparison of calculated spectra with experimental data is shown in \cref{fig:MTSAuAu}.
The normalization of the theoretical curves is chosen such that the overall multiplicities agree with the experimental ones.
The calculated and the experimental spectra for the $K^+$ agree prominently well. The comparison for the $K^-$ (and even more for the $\phi$ mesons) is complicated by the statistical fluctuations, both of experimental data and calculated results.

The resulting spectra from the GiBUU default scenario, cf.~\cref{fig:MTSAuAuHadron}, show up to be different: the slope parameter of the $K^+$ is larger ($\sim 20\MeV$), while the slope of the $K^-$ is smaller ($\sim40\MeV$). As explained above, spectra looking quite similar can lead to fitted slope parameters differingy by tens of \MeV.

\section{Conclusions}
\label{sec:Conclusions}

In the present paper, a consistent way of including heavy resonances via Hagedorn resonances into the hadronic transport model GiBUU is presented and used for the first time for a full dynamical microscopical calculations of heavy ion collisions.
\HSS implement multi-baryonic (and highly unstable) states as 'energy reservoirs'. The resonances are constructed according a 'microcanonical' bootstrap with explicit conservation of baryon number, strangeness and isospin. One single parameter, the radius of one such \HS, governs the full prescription. It is chosen such, that the mass degeneration spectra exhibit an exponential slope with a slope parameter $T_H\simeq 165\MeV$. In addition, this parameter also fixes the production cross section of \HSS in $2\to1$ processes and, via detailed balance, also the decay width for the decays $1\to2$.
These processes replace the conventional hadronic interactions and thus present a new unorthodox picture for production and absorption of hadrons.

In order to compare this scenario with experimental multiplicities of charged particles in nucleon-nucleon collisions, a strangeness suppression factor $\gamma_s$ has to be induced. Here a good overall agreement can be achieved.

This transport model is applied to A+A collisions in the region of $\sqrt{s_{NN}}\simeq 2.5\GeV$ as measured by the HADES collaboration. The resulting multiplicities of pions and strange hadrons are in good agreement. Special consideration has to be done for the $\phi$ meson. Here the yields comes out too large. Nevertheless, one observes that the production of $K^-$ mesons is mainly governed by the decay of the $\phi$ meson, which can be quantified by a $\phi/K^-\simeq 0.8$ ratio. This is higher than the experimental observed value.

The slopes of the $m_T$-spectra are nicely reproduced for $K^+$, $K^-$ and $\phi$ mesons. The contribution of $\phi$ decays to the $K^-$ spectra is important.

In order to test theses findings, also calculations with the default hadronic treatment of the GiBUU transport code are performed. The resulting multiplicities are nearly identical to those within the Hagedorn treatment. Again, the $\phi$ meson plays a special role. In the traditional hadronic picture, the yield of $\phi$ meson is clearly underestimated. Nevertheless, since the $K^-$ yields are the same, one clearly observes, that in this picture, the production mechanism of the $K^-$ meson is different. The experimental $\phi/K^-$ is underestimated.

Also the slopes of the $K^+$ and the $K^-$ are not so well reproduced. While the $K^+$ come out steeper than in the Hagedorn picture, the $K^-$ slope parameter is larger.

It would thus be instructive to study in a next step the thermodynamical evolution of the collision system in both prescriptions and check the degree of thermalization reached. Temperature profiles in space and time could be easily extracted by reporting the actual values of the energy momentum tensor. This is left for future studies.

It has to be checked, whether a larger radius of the \HS, yielding a lower Hagedorn temperature $T_H$, would also yield lower slope parameters of the spectra in these A+A collisions. This could be connected then with the averaged number of pions in N+$\overline{\textrm{N}}$ annihilations at rest, which comes a little bit too low in the present prescription at the moment. All these studies will be necessary for tests of the applicability of the Hagedorn picture to systems with vanishing baryo-chemical potential as e.g.~the matter produced in ultra-relativistic heavy ion collisions.

A statistical bootstrap model, embedded into a (hadronic) transport model is thus a valuable tool to study not only very energetic, ultrarelativistic heavy ion collisions, but also the behavior in such collisions at low energies. By providing a total different production mechanism for strange particles it supplements thus a usual treatment and thus provides new insights into the properties of the produced matter in these collisions.

An extension of the given picture would allow in the field of ultra-relativistic heavy ion collisions for the study of production of heavier quark states as e.g.~charm or also the production of light nuclei like deuterons, tritons etc.. Furthermore, with a bootstrap picture of color neutral heavy states from the partonic side, a real microscopic picture of the phase transition of a partonic system to a pure hadronic system via Hagedorn resonances could be developed.

\begin{acknowledgments}
The authors thank H.~Schuldes and M.~Lorenz for very helpful discussions concerning the HADES data and providing access to it.

This work was supported by the Bundesministerium f\"ur Bildung und
Forschung (BMBF). Numerical computations have been performed at the
Center for Scientific Computing (CSC).
\end{acknowledgments}

\bibliography{Complete,Literature}

\end{document}